# High precision framework for Chaos Many-Body Engine

I.V. Grossu[a,*], C. Besliu[a], D. Felea[b], Al. Jipa[a]

[a] *University of Bucharest, Faculty of Physics, Bucharest-Magurele, P.O. Box MG 11, 077125, Romania*
[b] *Institute of Space Science, Laboratory of High Energy, Astrophysics and Advanced Technologies, Bucharest-Magurele, P.O. Box MG 23, 077125, Romania*

ABSTRACT

In this paper we present a C# 4.0 high precision framework for simulation of relativistic many-body systems. In order to benefit from the, previously developed, chaos analysis instruments, all new modules were integrated with Chaos Many-Body Engine [1,2]. As a direct application, we used 46 digits precision for analyzing the "Butterfly Effect" of the gravitational force in a specific relativistic nuclear collision toy-model.

**1. Introduction**

In [1] we presented Chaos Many-Body Engine, a C# 2.0 application for chaos analysis of relativistic many-body systems. Although the double value type (8 octets, 15-16 digits precision) is sufficient for a wide class of simulations, it involves unacceptable limitations [3] in some specific problems encountered in physics (e.g. in the natural system of unities, the magnitude order of the gravitational constant is $10^{-45}$ $MeV^{-2}$). Considering the high instability on fluctuations and perturbations of non-linear systems, neglecting such small quantities might result in important lose of information. In this context, we considered the opportunity of developing a high precision version of Chaos Many-Body Engine.

**2. Program description**

The "Chaos Many-Body Engine" library is designed as a flexible, general numerical solution for the simulation of three-dimensional relativistic many-body systems. In order to benefit from the latest technologies advantages, the code was first migrated to .Net Framework 4.0 [2].

**Math.dll**. The *BigInteger* structure is a new feature in .Net Framework 4.0 [4]. For big rational numbers, Microsoft provides also an additional open source structure (*BigRational*), which is designed as a two *BigInteger-s* ratio.

Trying to use *BigRational*, one first problem comes from a growing precision effect (involved especially by multiplications), which leads to an unacceptable performance deprecation. Thus, starting from *BigRational*, we created the *BigDecimal* structure, its precision being limited (the *LimitPrecision* method) to the number of digits specified by the *Decimals* static property.

Another difficulty comes from the fact that .Net Framework 4.0 does not provide common mathematical functions for *BigRational-s*. In this context, we developed the static *BigMath* class, which implements a *halving interval programming technique* for computing the square root and the exponential functions for *BigDecimal* numbers.

The *BigVector* class is used for abstracting high precision three-dimensional vectors. We used operator overloading for implementing the sum, scalar product, and vectorial product operations.

*Corresponding author.
E-mail addresses: *ioan.grossu@brahms.fizica.unibuc.ro*(I.V.Grossu)

**AutomaticTests.dll**. As Visual C# 2010 Express Edition does not include unit test capabilities, we created a simple framework for supporting automatic tests.

The *AutomaticTest* attribute represents the metadata used for identifying test methods. The *BaseAutomaticTest* class is designed as the base class for all automatic tests classes. The *AutomaticTestsEngine* contains a generic list of *BaseAutomaticTest* objects, and uses *reflection* [4] for dynamically calling all corresponding test methods (methods with the, previous mentioned, *AutomaticTest* attribute).

Detailed automatic tests were implemented for the most important classes of the Math.dll library (*BigDecimal, BigMath, BigVector, BigRelativity*), in order to minimize the impact of further changes or refactoring processes.

**ParallelProgramming.dll**. As working with high precision numbers might seriously affect performance, an important attention was paid to both code optimization and algorithm parallelization. The *ParallelProgramming* library is a basic framework for executing methods on different execution threads [4,5]. The *IParallelObject* interface is used for defining the "contract" (*Start* and *Stop* methods) that a class must conform in order to be parallelizable. The *ParallelTask* class is used to create a new thread for executing the *Start* method of an *IParallelObject* instance, received as a constructor parameter (dependency injection). The *ParallelTaskList* class is based on a generic list of *ParallelTask* objects, and is used for calling multiple methods in parallel. Further analyses along parallelization techniques are currently in progress.

**Engine.dll**. The high precision many-body library was developed by analogy with the existing many-body engine [1,2]. Thus, the *BigParticle* class abstracts a material point and provides some basic scalar and vectorial properties (rest mass, movement mass, electric charge, velocity, force etc.). *BigNBody* is a generic list of *BigParticle* instances and provides also the main simulation algorithms. The *BigNBody.Next* method uses a second order Runge-Kutta algorithm [3] for computing the following system of equations (derived from the second Newton's law):

$$\begin{cases} \dfrac{d\vec{p}_i}{dt} = \sum_{i \ne j}' \vec{F}_{ij} = \vec{F}_i \\ m_i \dfrac{d\vec{r}_i}{dt} = \vec{p}_i \\ m_i = \dfrac{m_{0i}}{\sqrt{1 - \left(\dfrac{v_i}{c}\right)^2}} \end{cases} \quad (1)$$

where $p_i$ is the momentum, $r_i$ the position, $v_i$ the velocity, $m_{0i}$ the rest mass, $m_i$ the movement mass of the constituent $i$, $t$ is the time, $F_{ij}$ the bi-particle force, and $c$ is the velocity of light in vacuum.

The *BigNBody.EnergyTest* get property is used for implementing the energy conservation assessment [1,3]:

$$P_e = -log_{10} \left| \dfrac{E(t) - E(t = 0)}{E(t = 0)} \right| \quad (2)$$

The expression of the bi-particle force is defined in each specific simulation of interest, and is passed to the engine using a delegate (type-safe function pointer). *BigUniverse* is a generic list of *BigNBody* instances, and was mainly used for implementing the Lyapunov Exponent [6,7] by simulating two identical systems with slightly different initial conditions. It is implementing the, previously described, *IParallelObject* interface, and contains also the simulation main loop (time generator). *BigSimulationBase* represents the base abstract class to be implemented by each specific simulation in order to enforce a proper use of the engine. It represents the application programming interface (*facade layer*) of the engine.

**Data.dll.** Each simulation output is saved into a set of comma separated values files: the header ([name].hdr.csv) containing the initially momentum and position for each particle, and the data file ([name].dat.csv) containing the system evolution in time information. One limitation of the current version comes from the fact that the output files are storing data in *decimal* precision (.Net data type with 28-29 digits) instead of using a specific *BigDecimal* representation (e.g. numerator/denominator). Considering also that the old CMBE version is using double precision, our choice is motivated by the fact that both the *decimal* and the *BigDecimal* data types are represented as base 10 numbers (the numerator and the denominator properties of *BigDecimal* structure are base 10 integers), while the *double* .Net data type is binary represented. Thus, in order to avoid any small precision loss resulted from converting *decimal* to *double* variables [4] we created also a *decimal* version of the data access layer module (*Data.dll*), which include all chaos analysis tools implemented in [1].

**Functional Tests.** As test strategy, we implemented a simple relativistic three-body system with harmonic potential (choose Simulations\Harmonic, respectively High Precision\Harmonic from menu) for comparing the new high precision module, configured to work in double precision (*BigDecimal.Decimals* = 16), with the old CMBE version. The same parameter values (temporal resolution, simulation time etc.) were provided to both simulations. The dependence on time of the phase space distance between the two outputs is presented in Fig.1. The small difference found could be explained by the previously discussed precision loss resulted from converting *decimals* to *doubles* (base 10 vs. base 2 number representations). For proving this assumption we created a *decimal* version of the old CMBE application (copy/paste all code files, and replace all *double* with *decimal* variables). In order to assure the compatibility with the *decimal* data type, the *BigDecimal.Decimals* static variable was increased to 29. One can notice that, in this case, the phase space distance between the two simulations is constant and equals zero. We observed also that the energy conservation test has approximately the same value ($P_\varepsilon \approx 6$) for all scenarios we tested (CMBE old, respectively the high precision module configured for 16, 29, and 46 decimals).

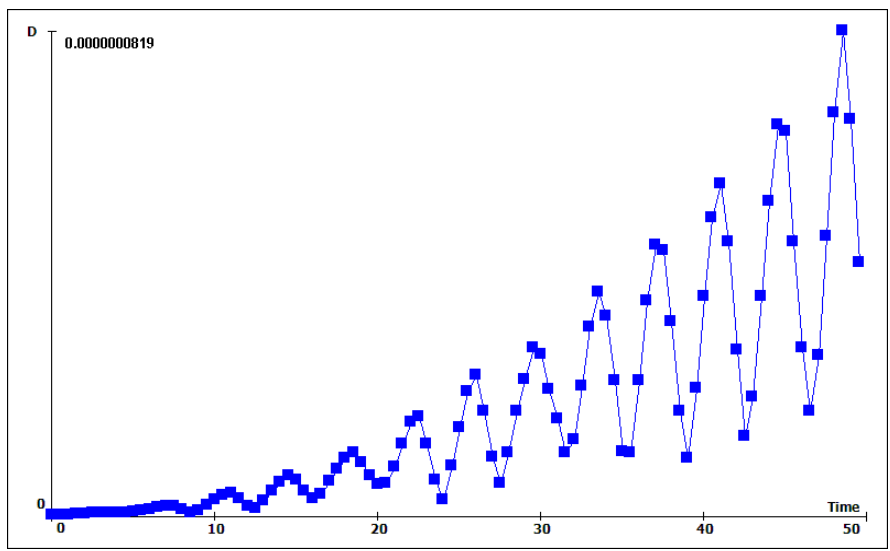

**Fig.1.** The phase space distance (*D*) between the high precision module (configured to 16 decimals precision) and the old CMBE simulation outputs for a simple relativistic three body system with harmonic potential.

**3. Application: Study on the "Butterfly Effect" of the gravitational force in a nuclear relativistic collision toy-model**

Inspired by existing studies on billiard nuclear models [8,9], we used Chaos Many-Body Engine for implementing a relativistic nuclear collision toy-model, which was tested with encouraging results [10] on experimental data from SKM 200 collaboration [11-13]. In the frame of this model we considered a bi-particle finite depth Yukawa potential well, together with a short distance repulsive term. As opposed to the potential implemented in [10], for simplicity, we ignored the coulombian term:

$$V(r_{ij}) = \begin{cases} -K\ln(r_{ij}), & r_{ij} \leq 1.134\,Fm \\ -V_0 \frac{e^{-r_{ij}/a}}{r_{ij}/a} - K\ln(r_{ij}), & r_{ij} \in (1.134, 2.2)\,Fm \\ -V_0 \frac{e^{-r_{ij}/a}}{r_{ij}/a}, & r_{ij} \geq 2.2\,Fm \end{cases} \quad (3)$$

where $V_0=35\,MeV$ is the depth and $a=2\,Fm$ is the radius of the potential well, $r_{ij}$ represents the distance between the two bodies, and we empirically chose $K=200$.

Each nucleus is implemented as a set of nucleons, initially at rest, placed in the vertices of a regular polyhedron (tetrahedron for α particle, cubical network for heavier nuclei), with edges calculated in agreement with the corresponding nucleus radius ($r=r_0 A^{1/3}$). The incident momentum per nucleon is specified as an input parameter. One can choose between *target-projectile*, and *collider* mechanisms. Reactions are ignored in order to avoid additional hazard involved by multiple channels.

In order to estimate the *Butterfly Effect* of the gravitational force, in each simulation we included a clone of the main many-body system, for which we considered an additional gravitational term:

$$V_G(r_{ij}) = V(r_{ij}) - G\frac{m_i m_j}{r_{ij}} \quad (4)$$

where *m* is the movement mass, and *G* the gravitational constant.

As in the natural system of unities $G \cong 6.7 \times 10^{-45}\,MeV^{-2}$, the precision was set to 46 digits. In order to exclude any computational bug, the code was first tested by ignoring the gravitational term. Thus, we verified that the phase space distance between the main and the cloned systems is constant and equals zero.

For exemplification, we considered He+He target-projectile central collisions at 4.5 AGeV/c (choose High Precision\Gravity from menu). With these parameters, the energy conservation test (2) is $P_\varepsilon \approx 4$ for the temporal resolution $dt=10^{-2}\,Fm/c$, respectively $P_\varepsilon \approx 5$ for $dt=10^{-3}\,Fm/c$ (same values being obtained with the old version of CMBE). The phase space distance between the two systems, with and without gravity, is presented in Fig.2.

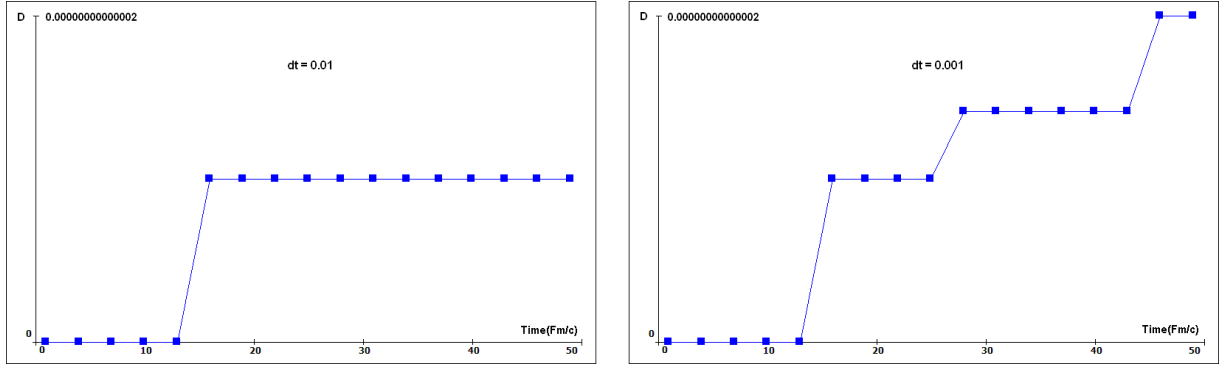

**Fig.2.** The phase space distance (*D*) between two many-body systems (He+He at 4.5 A GeV/c), one with and one without gravity (4). (Left) The result for temporal resolution *dt = 0.01 Fm/c*. (Right) Temporal resolution *dt = 0.001 Fm/c*.

### 4.Conclusion

Based on the *BigInteger* and *BigRational* .Net Framework 4.0 new features, we developed a C# [4,5] high precision [3] relativistic many-body framework, fully integrated with Chaos Many-Body Engine [1,2]. Working with 46 digits precision involves important performance limitations. A dual core @ 2.0 GHz CPU is a minimum recommendation in this context. The number of participants was also limited to 24 (e.g. C + C interactions).

Trying to estimate the *Butterfly Effect* of the gravitational force in a specific relativistic nuclear collision toy-model, we obtained some encouraging results. Thus, despite the very small value of the gravitational term (G $\cong$ 6.7×10$^{-45}$ MeV$^{-2}$), one can notice that its contribution results in phase space distances "visible" in double precision (16 decimals) Fig.2.

**Acknowledgement:** This study is supported by a grant of the Romanian National Authority for Scientific Research, CNCS – UEFISCDI, project number 34/05.10.2011 PN-II-ID-PCE.

### References

[1] I.V. Grossu, C. Besliu, Al. Jipa, C.C. Bordeianu, D. Felea, E. Stan, T. Esanu, Comput.Phys. Comm. 181 (2010) 1464–1470.84.
[2] I.V. Grossu, C. Besliu, Al. Jipa, D. Felea, T. Esanu, E. Stan, C.C. Bordeianu, Comput.Phys. Comm. 184 (2013) 1346-1347
[3] R.H. Landau, M.J. Paez, C.C. Bordeianu, Computational Physics: Problem Solving with Computers, Wiley-VCH-Verlag, Weinheim, 2007, pp. 13-24, 215-225
[4] Joseph Albahari, Ben Albahari, C# 4.0 in a Nutshell, O'Reilly, US, 2010, pp. 27, 237-240, 681-733, 789-941
[5] A. Ignat, Asp.NET MVC Tips and Tricks, CreateSpace, US, 2009
[6] M. Sandri, Numerical Calculation of Lyapunov Exponents, Mathematica J.6, 78-84, 1996.
[7] C.C. Bordeianu, D. Felea, C. Besliu, Al. Jipa, I.V. Grossu, Comput. Phys. Comm. 178 (2008) 788–793.

[8] G.F. Burgio, F.M. Baldo, A. Rapisarda, P. Schuck, Phys. Rev. C 58 (1998) 2821-2830
[9] D. Felea, C.C. Bordeianu, I.V. Grossu, C. Besliu, Al. Jipa, A.-A. Radu, E.Stan, EPL, 93 (2011) 42001
[10] I.V. Grossu, C. Besliu, Al. Jipa, E. Stan, T. Esanu, D. Felea , C.C. Bordeianu, Comput.Phys. Comm. 183 (2012) 1055-1059.
[11] Al. Jipa, C. Besliu, Elemente de .zica nucleara relativista – Note de curs, Editura Universitatii din Bucuresti, Bucuresti, Romania, 2002.
[12] A.U. Abdurakhimov, et al., preprint JINR 13-10692, 1977.
[13] C. Besliu, N. Ghiordanescu, M. Pentia, Studii si Cercetari de Fizica 29 (1977) 817.